# Synchronization of two coupled multimode oscillators with time-delayed feedback

Yulia P. Emelianova[a,b,*], Valeriy V. Emelyanov[b], Nikita M. Ryskin[b]

[a] *Institute of Electronics and Mechanical Engineering, Yuri Gagarin State Technical University of Saratov, Polytechnicheskaya 77, Saratov 410054, Russia*

[b] *Department of Nonlinear Processes, Saratov State University, Astrakhanskaya 83, Saratov, 410012, Russia*

**Abstract**

Effects of synchronization in a system of two coupled oscillators with time-delayed feedback are investigated. Phase space of a system with time delay is infinite-dimensional. Thus, the picture of synchronization in such systems acquires many new features not inherent to finite-dimensional ones. A picture of oscillation modes in cases of identical and non-identical coupled oscillators is studied in detail. Periodical structure of amplitude death and "broadband synchronization" zones is investigated. Such a behavior occurs due to the resonances between different modes of the infinite-dimensional system with time delay.

## 1. Introduction

Study of synchronization effects in systems of coupled oscillators is an important problem of the modern nonlinear dynamics [1-5]. In particular, a problem of synchronization of time-delayed systems is of great importance because such systems are widespread in neuronal dynamics, nonlinear optics, biophysics, geophysics, telecommunication and information engineering, economics, and ecology [3,6-21]. Systems with time delay are usually described by functional delay-differential equations (DDEs). DDEs are known to have infinite-dimensional phase space [17,18] and are capable of demonstrating a variety of dynamic regimes including chaos [3,9-12,15].

Among the various effects which are observed in systems of interacting oscillators, the amplitude death (AD) effect has been a topic of interest [1,22-29]. AD is a phenomenon of oscillation suppression as a consequence of dissipative coupling. Such a behavior is also characterized by suppression of amplitudes to zero values. AD is desirable in various applications where fluctuations should be suppressed and a constant output is needed. In particular, the amplitude death caused by the time delay in coupling has been studied in many works [25-29]. On the contrary, in [30] it was shown that the delay coupling may induce chaotic behavior, even in a simple system of coupled oscillators.

Features of synchronization of time-delayed systems have been studied in several works. A comprehensive review of the problem has been recently given in [21]. In particular, Usacheva and Ryskin [31] have investigated the forced synchronization of a delayed-feedback oscillator driven by an external harmonic signal. Mensour and Longtin [32] have considered drive-response synchronization of two time-delayed systems with application to secure communication. Ghosh *et al.* [33] have investigated a problem of synchronization between two forced oscillators with unidirectional coupling, when one oscillator has the time delay. In [34], Ghosh *et al.* have studied a design of delay coupling for targeting desired regime (synchronization, anti-synchronization, lag-synchronization, amplitude death, and generalized synchronization) in mismatched time-delayed dynamical systems.

---

[*] Corresponding author at: Institute of Electronics and Mechanical Engineering, Yuri Gagarin State Technical University of Saratov, Polytechnicheskaya 77, Saratov 410054, Russia. Tel.: +7 9172021488.
E-mail addresses: yuliaem@gmail.com (Y.P. Emelianova), emvaleriy@gmail.com (V.V. Emelyanov), ryskinnm@info.sgu.ru (N.M. Ryskin).

There exists a great variety of coupling topologies between oscillators with delay. In particular, the networks with time-delayed dissipative linear coupling, as well as pulse-coupled time-delayed networks, have been widely investigated [21]. Many of radio-frequency, microwave, and optical oscillators utilize a power amplifier with input connected with output via a time-delayed feedback loop [11,12,31,35,36]. The most natural way to couple two of such systems is to feed a portion of power from the feedback loop of one oscillator into the feedback loop of another one, and vice versa. In such a case, we have a specific type of coupling, i.e. a nonlinear time-delayed dissipative coupling, which has not been studied previously. In Sec. 2, we consider a model of coupled oscillators described by a system of coupled DDEs. In Sec. 3, we study synchronization of two oscillators with identical parameters. Sec. 4 contains extension to the case of non-identical oscillators. Oscillators with frequency mismatch, as well as with non-identical excitation parameters which determine the oscillation amplitudes, are considered. Transitions between different synchronization regimes are investigated. A special attention is paid to peculiarities of amplitude death and "broadband synchronization" (BS). BS has previously been observed in ensembles of finite-dimensional dissipatively coupled oscillators with non-identical excitation parameters, which are responsible for the oscillation amplitudes [4,37,38]. In such systems there appears a domain of synchronous regimes, which looks like a narrow band located between the AD and quasi-periodic domains and extends to very large values of the frequency mismatch. In the BS domain, the oscillator with larger amplitude dominates and suppresses natural oscillations of the other oscillators.

## 2. Model and basic equations

### 2.1. Single delayed-feedback oscillator

Consider a general scheme of a delayed-feedback ring-loop oscillator presented in Fig. 1(a). The oscillator consists of a nonlinear power amplifier, a bandpass filter, and a feedback leg containing a delay line, a variable attenuator, and a phase shifter. The filter is assumed to be narrow-band with Lorenz-shape frequency response. In such a case, it is convenient to express the signal as a quasi-harmonic oscillation with slowly varying amplitude $A(t)$ and carrier frequency $\omega_c$: $A(t)\exp(i\omega_c t)$. It is convenient to choose $\omega_c = \omega_0$ where $\omega_0$ is a central frequency of the filter passband (Fig. 1(a)). Under this assumption, one can derive an equation describing the dynamics of the slow amplitude [11,12,31]

$$\frac{dA}{dt} + \gamma A - \alpha f\left(\left|A(t-\tau)\right|\right)e^{i\left[\Phi\left(\left|A(t-\tau)\right|\right)+\theta\right]} = 0. \qquad (1)$$

Here $\gamma = \omega_0/2Q$ is the parameter of losses, $Q$ is the filter Q-factor, $\alpha = \gamma\rho G$ is the parameter of excitation, $\rho$ is the amount of feedback, $G$ is the small-signal gain factor of the amplifier, $\theta$ is the phase shift in the feedback loop, $f$ and $\Phi$ are nonlinear amplitude and phase transfer functions of the amplifier, respectively, and $\tau$ is the delay time.

Further we suppose that the nonlinear amplitude response of the amplifier is approximated by a cubic polynomial and neglect the phase nonlinearity. In such a case, Eq. (1) becomes

$$\frac{dA}{dt} + \gamma A - \alpha e^{i\theta}\left(1 - \left|A(t-\tau)\right|^2\right)A(t-\tau) = 0. \qquad (2)$$

Note that without the delay Eq. (2) becomes the well-known normal form for Andronov–Hopf bifurcation [39].

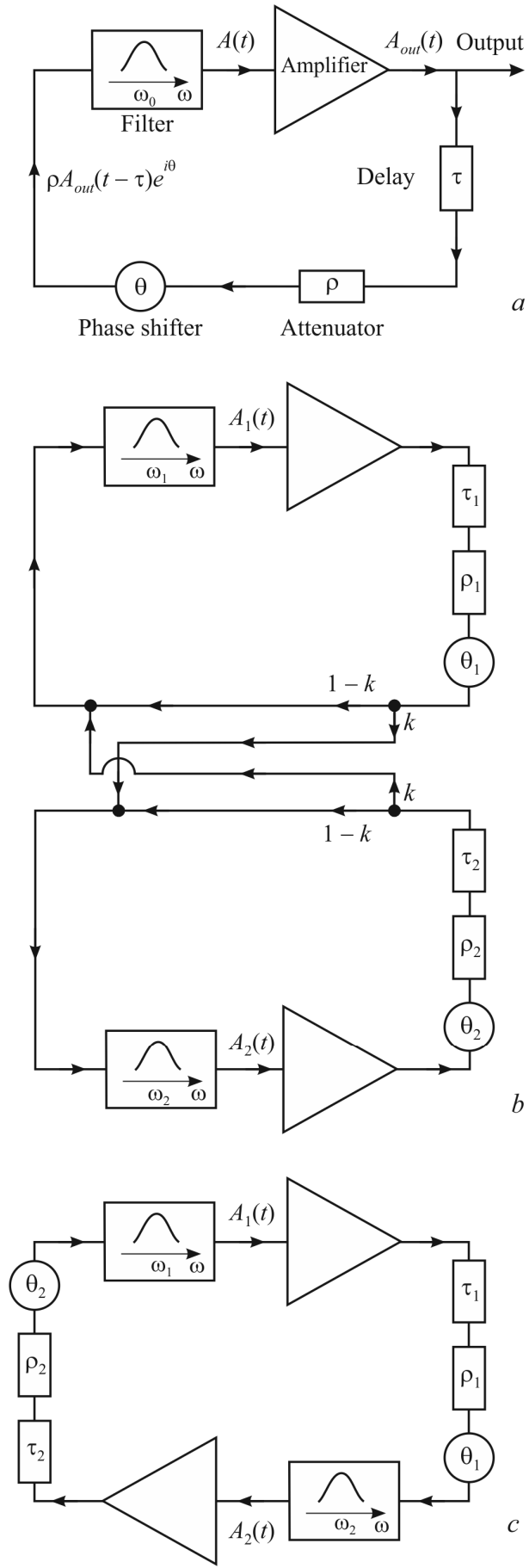

**Fig. 1.** Schematic diagrams (a) of a delayed-feedback oscillator, (b) of two coupled oscillators, and (c) of a two-stage oscillator.

Dynamics of this oscillator has been studied in detail [31,35,36]. It was shown that for a single-frequency solution $A = R_0 \exp(i\omega t)$, where $R_0$ may assumed to be real without loss of generality, the eigenfrequencies obey the equation

$$\omega = -\gamma \tan(\omega\tau - \theta). \tag{3}$$

This equation has infinite number of complex roots, since time-delayed systems with infinite-dimensional phase space have infinite number of eigenmodes. However, only the roots with $\cos(\omega\tau - \theta) > 0$ correspond to stable eigenmodes. Their amplitudes satisfy the equation

$$R_0^2 = 1 - \frac{\sqrt{\gamma^2 + \omega^2}}{\alpha} = 1 - \frac{\alpha_{st}}{\alpha}. \tag{4}$$

These solutions exist when the excitation parameter $\alpha$ exceeds the self-excitation threshold

$$\alpha_{st} = \sqrt{\gamma^2 + \omega^2}. \tag{5}$$

The self-excitation boundary on the $\theta-\alpha$ parameter plane has the shape of a periodic set of domains named "generation zones". In the centers of such generation zones at $\theta = 2\pi n$ the frequency $\omega = 0$, i.e., generation arises with frequency equal exactly to the resonance frequency of the filter. Accordingly, the self-excitation threshold is minimal: $\alpha_{st} = \gamma$. At the borders of generation zones at $\theta = 2\pi n + \pi$ two eigenfrequencies are spaced equally from the central frequency. Here, there are domains of bistability where simultaneous self-excitation of two modes is possible. Such a picture is common for many delayed-feedback oscillators [12,35,36,40].

If $\alpha$ exceeds a certain threshold $\alpha_{sm}$ the single-frequency oscillation becomes unstable via the Andronov–Hopf bifurcation. As a result, multiple-frequency self-modulation regime arises. In the center of a generation zone ($\theta = 2\pi n$) the self-modulation threshold is described by a simple formula [35]

$$\alpha_{sm} = \frac{1}{2}\left(3\gamma + \sqrt{\gamma^2 + \omega_{sm}^2}\right). \tag{6}$$

Self-modulation frequency $\omega_{sm}$ is a root of (3) with $\cos(\omega\tau - \theta) < 0$. Despite the fact that modes with $\cos(\omega\tau - \theta) < 0$ are unstable, they play an important role, because self-modulation is caused by their excitation on the background of the fundamental mode with high amplitude [35,36]. Further we refer to them as self-modulation modes. The self-modulation mode with frequency nearest to the center of the passband has the lowest threshold.

Further increase in $\alpha$ above the self-modulation threshold results in transition to chaos. The most common scenario is transition through a sequence of self-modulation period doubling bifurcations [35,36].

*2.2. Two oscillators with nonlinear time-delayed dissipative coupling*

The most natural way to couple two delayed-feedback oscillators is to feed a power portion from the feedback loop of one oscillator into the feedback loop of another one. Schematic diagram of two coupled oscillators is shown in Fig. 1(b). Similar type of coupling has been previously considered in application to synchronization of surface acoustic wave delay-line sensors [41]. However, in [41] the bifurcation analysis has been performed by reducing the infinite-dimensional DDE system to a system of ordinary differential equations on a center manifold of a rest point. Synchronization of two vacuum-tube microwave klystron oscillators with delayed feedback has been experimentally demonstrated in [42].

Let $k$ be a parameter defining the coupling strength as shown in Fig. 1(b). We can derive the following system of equations for the slowly varying amplitudes

$$\frac{dA_1}{dt} + \frac{i\Delta}{2}A_1 + \gamma_1 A_1 = \alpha_1 e^{i\theta_1}\left[(1-k)\left(1-|A_1(t-\tau_1)|^2\right)A_1(t-\tau_1) + k\left(1-|A_2(t-\tau_2)|^2\right)A_2(t-\tau_2)\right],$$

$$\frac{dA_2}{dt} - \frac{i\Delta}{2}A_2 + \gamma_2 A_2 = \alpha_2 e^{i\theta_2}\left[(1-k)\left(1-|A_2(t-\tau_2)|^2\right)A_2(t-\tau_2) + k\left(1-|A_1(t-\tau_1)|^2\right)A_1(t-\tau_1)\right]. \tag{7}$$

Here $\Delta = \omega_1 - \omega_2$ is the frequency mismatch, $\omega_{1,2}$ are central frequencies of the oscillators. The carrier frequency now is $\omega_c = (\omega_1 + \omega_2)/2$. Thus, we obtain a system of coupled oscillators with a specific type of coupling, i.e. a *nonlinear time-delayed dissipative coupling* which has not been studied previously. Consequently, one can expect the phenomena typical for systems with dissipative coupling, such as amplitude death [1,22-28] and "broadband synchronization" [4,37,38]. In this study, we focus on the structure of synchronization picture for identical and non-identical coupled oscillators with nonlinear time-delayed dissipative coupling (7). We obtain analytical boundaries for the domain of broadband synchronization and investigate transition mechanisms between different synchronization modes.

*2.3. Two-stage oscillator*

For better understanding of synchronization picture of (7) it is useful to consider the limit $k \to 1$ when the system of two coupled oscillators turns into a two-stage oscillator, which consists of two amplifiers chained in a ring loop (Fig. 1(c)). Recently, a two-stage klystron oscillator has attracted attention as compact powerful source of millimeter-wavelength radiation [43,44].

Let us briefly describe the pattern of oscillation regimes taking place with gradual increase in the excitation parameter while keeping other parameters constant, i.e., assuming that the gain of the amplifiers increases slowly. For simplicity, consider two amplifiers with identical parameters. Eqs. (7) become

$$\frac{dA_1}{dt} + \gamma A_1 = \alpha e^{i\theta}\left(1 - |A_2(t-\tau)|^2\right)A_2(t-\tau),$$
$$\frac{dA_2}{dt} + \gamma A_2 = \alpha e^{i\theta}\left(1 - |A_1(t-\tau)|^2\right)A_1(t-\tau). \quad (8)$$

When $\alpha$ exceeds the self-excitation threshold $\alpha_{st}$, which coincides with that of the single-stage oscillator (Eq. (5)), single-frequency mode of oscillation is excited. Just above the threshold, the oscillation is symmetric: $|A_{1,2}| = R_0$. Frequency and amplitude of this oscillation are defined by Eqs. (3), (4), i.e., they are the same as for the single-stage oscillator.

When $\alpha > 2\alpha_{st}$, a symmetry breaking takes place and the amplitudes $R_{1,2} = |A_{1,2}|$ become unequal

$$R_\pm^2 = \frac{1}{2}\left(1 \pm \sqrt{1 - (2\alpha_{st}/\alpha)^2}\right). \quad (9)$$

Note that there exist two possible solutions, one with $R_1 = R_+$, $R_2 = R_-$, and the other one with $R_2 = R_+$, $R_1 = R_-$.

In Fig. 2, waveforms of the amplitudes $R_{1,2}$, phase plots, and spectra illustrating dynamics of the two-stage oscillator are presented. The spectra are plotted only for one of the signals, $A_2(t)$, the spectra for $A_1(t)$ look similar. Fig. 2(a) demonstrates the symmetry breaking bifurcation. When $\alpha$ exceeds the self-modulation threshold $\alpha_{sm}$, the single-frequency regime becomes unstable, and output signals begin to oscillate around the steady-state values. In the spectrum, pairs of satellites equally spaced from the fundamental frequency arise (Fig. 2(b)).

Further increase in $\alpha$ leads to a period-doubling transition to chaos (Figs. 2(c)-(e)). Well above the self-excitation threshold, two asymmetric attractors merge into one fully developed chaotic attractor. The waveforms become strongly irregular and spectra become more uniform (Fig. 2(e)). Such a behavior is in good agreement with the picture described in [43,44] for the two-stage klystron oscillator.

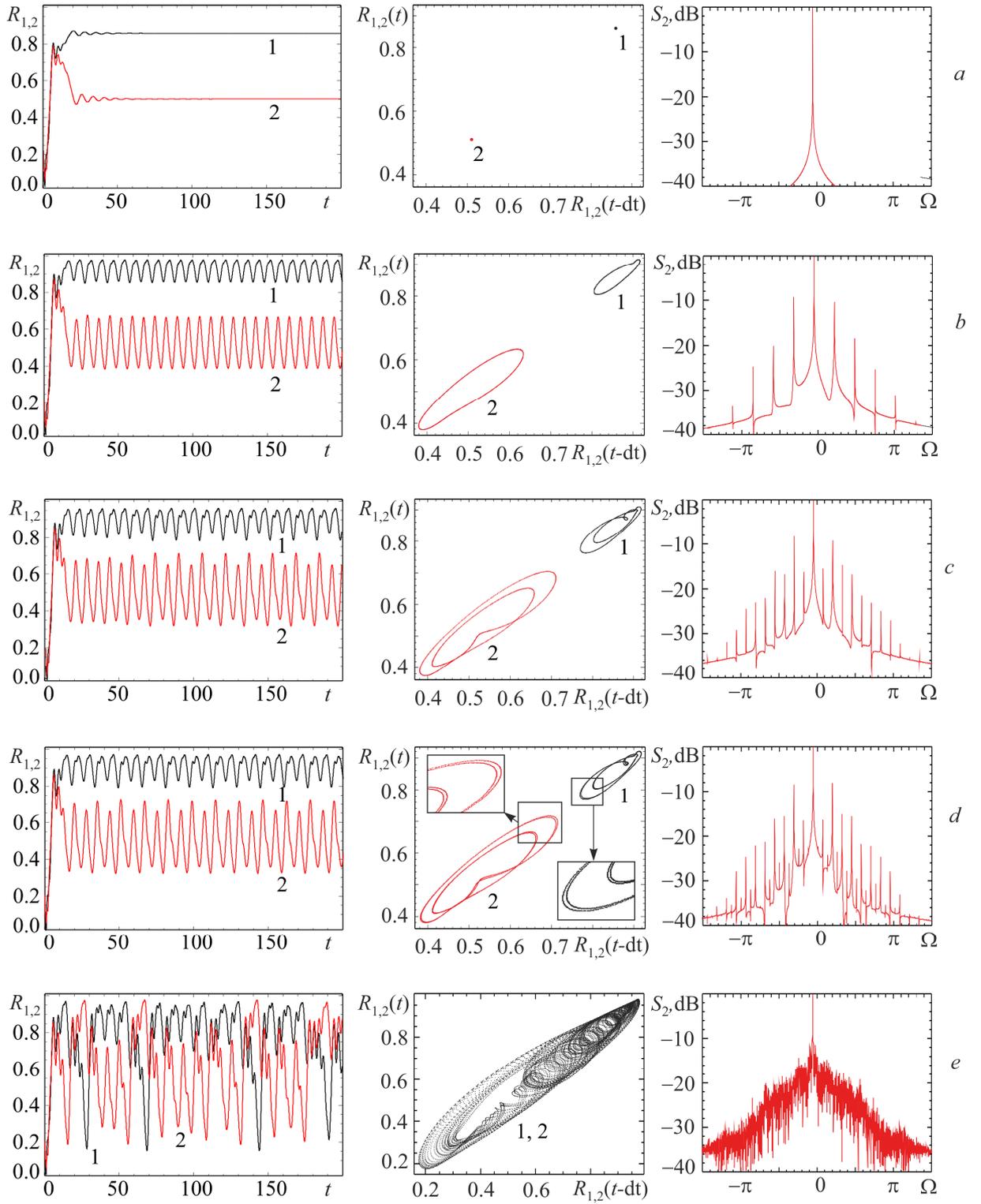

**Fig. 2.** Waveforms (left), phase portraits (center) and spectra (right) demonstrating symmetry breaking and transition to chaos for the two-stage oscillator (8). Values of the parameters are $\tau = 1.0$, $\gamma = 1.0$, $\theta = 0.1\pi$, (a) $\alpha = 2.3$, (b) 2.43, (c) 2.48, (d) 2.484, (e) 2.6.

## 3. Synchronization of two identical oscillators

### 3.1. Condition of symmetry breaking bifurcation

Let us start from study of synchronization of two coupled oscillators with identical parameters. Eqs. (7) become

$$\frac{dA_1}{dt} + \gamma A_1 = \alpha e^{i\theta}\left[(1-k)\left(1-|A_1(t-\tau)|^2\right)A_1(t-\tau) + k\left(1-|A_2(t-\tau)|^2\right)A_2(t-\tau)\right],$$
$$\frac{dA_2}{dt} + \gamma A_2 = \alpha e^{i\theta}\left[(1-k)\left(1-|A_2(t-\tau)|^2\right)A_2(t-\tau) + k\left(1-|A_1(t-\tau)|^2\right)A_1(t-\tau)\right]. \quad (10)$$

Evidently, there exists a regime of *complete synchronization*. In this regime, the amplitudes of both oscillators are the same, $A_1 = A_2$, and Eqs. (7) are reduced to Eq. (2) which describes a single oscillator. Also, there exist non-synchronous regimes with $A_1 \neq A_2$. Stability condition for the mode of complete synchronization can be derived in the same way as the symmetry breaking condition for the two-stage oscillator (Sec. 2.3). Seeking for steady state single-frequency solutions $A_{1,2} = R_{1,2}\exp(i\omega t)$, where $R_{1,2}$ may assumed to be real, from Eqs. (10) we obtain four real equations

$$\omega R_{1,2} = \alpha \sin(\theta - \omega\tau)\left[(1-k)\left(1-R_{1,2}^2\right)R_{1,2} + k\left(1-R_{2,1}^2\right)R_{2,1}\right], \quad (11)$$

$$\gamma R_{1,2} = \alpha \cos(\theta - \omega\tau)\left[(1-k)\left(1-R_{1,2}^2\right)R_{1,2} + k\left(1-R_{2,1}^2\right)R_{2,1}\right]. \quad (12)$$

From Eqs. (11), (12) we see that $\omega$ satisfies Eq. (3). Also, it is useful to rewrite Eqs. (11), (12) as follows

$$R_{1,2} = \frac{\alpha}{\alpha_{st}}\left[(1-k)\left(1-R_{1,2}^2\right)R_{1,2} + k\left(1-R_{2,1}^2\right)R_{2,1}\right], \quad (13)$$

where $\alpha_{st}$ is the self-excitation threshold given by Eq. (5). After some algebraic manipulations with Eq. (13), we find the following connection between $R_1$ and $R_2$:

$$(1-k)\left(R_1^2 - R_2^2\right)R_1 R_2 + k\left(R_1^2 - R_2^2 - R_1^4 + R_2^4\right) = 0. \quad (14)$$

From Eq. (14) it follows that there exists either symmetric solution ($R_1 = R_2$) or non-symmetric one, for which

$$(1-k)R_1 R_2 + k\left(1 - R_1^2 - R_2^2\right) = 0. \quad (15)$$

Because the non-symmetric solution branches off from the symmetric one, at the bifurcation point we have $R_1 = R_2 = R_0$ where $R_0$ obeys Eq. (4). Thus, substitution of Eq. (4) into Eq. (15) gives us the condition of symmetry breaking

$$\alpha_{sb} = \alpha_{st}\left(1 + \frac{k}{2k-1}\right). \quad (16)$$

In the limit $k \to 1$ from Eq. (16) we obtain the condition of symmetry breaking for the two-stage oscillator (Sec. 2.3) $\alpha_{sb} = 2\alpha_{st}$.

*3.2. Numerical results*

Let us study dynamics of the system of two coupled identical oscillators (10) when the phases are chosen close to the center of a generation zone ($\theta = 0.1\pi$). Numerical integration of Eqs. (10) was performed by the 4-order Runge–Kutta method adapted for DDEs [45]. Fig. 3 shows the chart of dynamical regimes on the $k - \alpha$ parameter plane, where different generation regimes are marked by different colors. If the excitation parameter $\alpha$ is less than the self-excitation threshold $\alpha_{st}$, the fixed point at the origin $A_1 = A_2 = 0$ is stable and oscillations of the both oscillators are damped. Above $\alpha_{st}$ there is a domain of completely synchronized steady-state generation with equal amplitudes. Above the self-modulation threshold $\alpha_{sm}$ there is a domain of non-steady-state complete synchronization. In this mode, the system demonstrates either quasi-periodic self-modulation or chaotic oscillation with equal instantaneous values of the amplitudes: $A_1(t) = A_2(t)$. Similar to the uncoupled oscillator (2), with an increase in $\alpha$ transition to chaos via the sequence of the self-modulation period doublings occurs. The boundary of transition to chaos is shown by the dashed line in Fig. 3. The period-doubling cascade occurs in a very short range of $\alpha$ values and is not shown in Fig. 3.

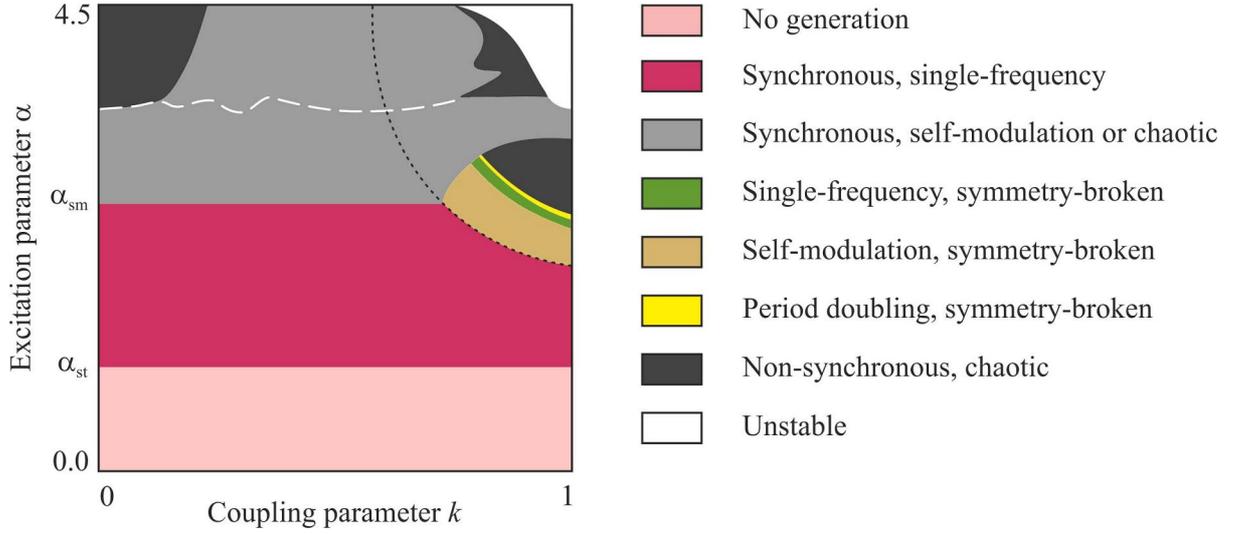

**Fig. 3.** Chart of dynamical regimes for the system (10) for $\tau = 1.0$, $\gamma = 1.0$, $\theta = 0.1\pi$. Dotted line shows analytical threshold of symmetry breaking (Eq. (16)). Dashed line indicates boundary of transition to synchronous chaos.

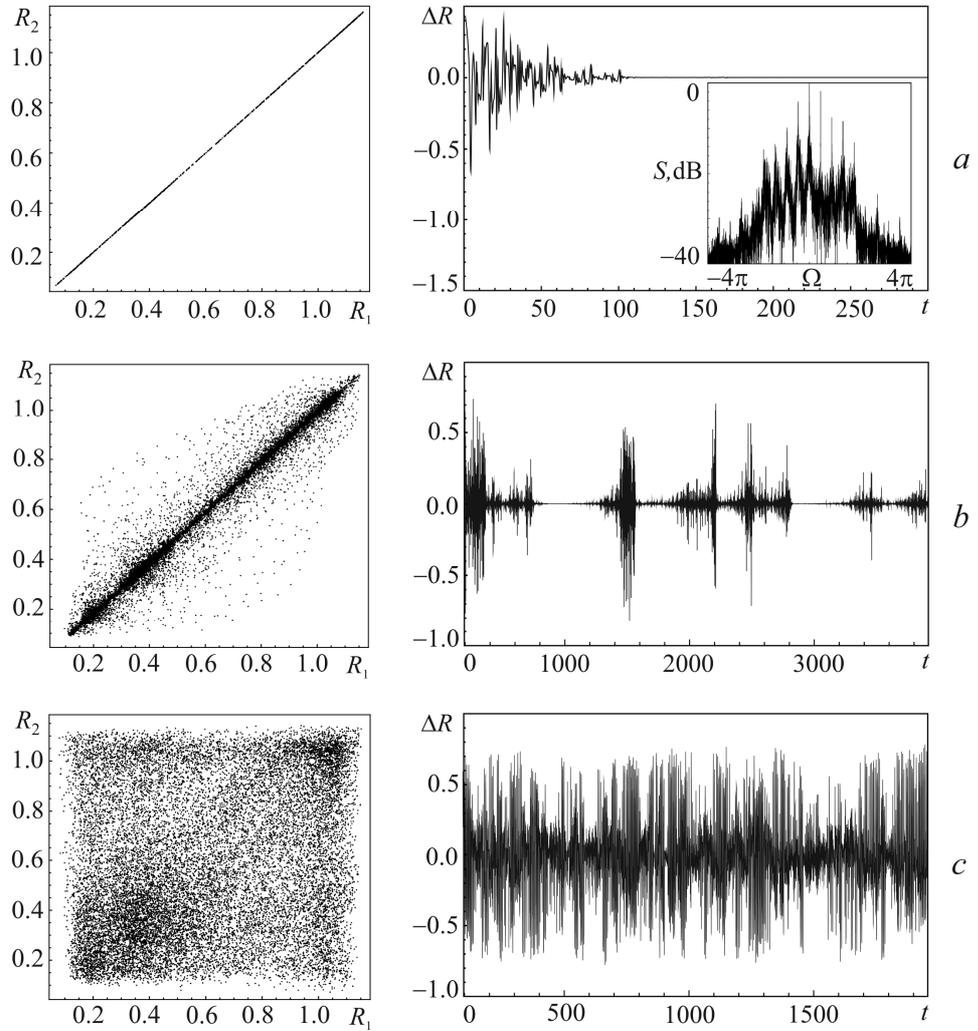

**Fig. 4.** Projections of the phase plots on the $R_1 - R_2$ plane (left) and waveforms of difference of the amplitudes (right) demonstrating transition from synchronous to asynchronous chaotic mode through the intermittency of "bubbling" type in the system (10) at $\alpha = 4.1$. The values of coupling parameters are (a) $k = 0.2$, (b) 0.16, and (c) 0.1. Other parameters are the same as in Fig. 3.

For small values of the coupling parameter $k$ the chaotic regime is asynchronous, i.e. $A_1(t) \neq A_2(t)$. Transition from synchronous to asynchronous generation occurs through the intermittency of "bubbling" type [46-48]. Fig. 4 shows projections of phase plots on the $R_1 - R_2$ plane and plots of amplitude difference $\Delta R = R_1 - R_2$ in time. One can see that with the decrease in $k$ we move towards the region of chaotic asynchronous generation. The duration of the "laminar" phase with $\Delta R \approx 0$ decreases. At the same time, the "turbulent" bursts become more frequent and their duration increases. For the regime of synchronous generation, projection of the phase plot looks like a straight line where $R_1 = R_2$ (Fig. 4(a)). Transition to the asynchronous regime corresponds to the perturbation of this straight line (Fig. 4(b)) which spreads so that the phase plot projection looks like the fully filled square (Fig. 4(c)).

At sufficiently high values of the coupling parameter $k$ the symmetry breaking bifurcation occurs when $\alpha$ exceeds the symmetry breaking threshold $\alpha_{sb}$ shown by the dotted line in Fig. 3. Above this line the steady-state mode with unequal values of the amplitudes is observed. With an increase in $\alpha$ self-modulation and transition to chaos via the period doubling cascade takes place, similar to the picture described in Sec. 2.3 for the two-stage oscillator (Fig. 2). The chaotic regime is asynchronous. With the decrease in $k$ transition to the completely synchronized mode occurs.

At high values of $\alpha$ and $k$ the solution becomes unstable and goes to infinity, similar to the time-delayed oscillator with cubic nonlinearity [35].

Note that the boundaries of self-excitation, self-modulation and symmetry breaking are in excellent agreement with theoretical predictions given by Eqs. (5), (6), and (16), respectively.

## 4. Synchronization of two non-identical oscillators

Consider synchronization of two non-identical oscillators. In this paper, we consider non-identity of the excitation parameters $\alpha_{1,2}$ and the frequency mismatch $\Delta$, which have the most significant impact on the oscillation amplitude and frequency. In that case, we rewrite Eqs. (7) as follows

$$\frac{dA_1}{dt} + \frac{i\Delta}{2}A_1 + \gamma A_1 = \alpha_1 e^{i\theta}\left[(1-k)\left(1-|A_1(t-\tau)|^2\right)A_1(t-\tau) + k\left(1-|A_2(t-\tau)|^2\right)A_2(t-\tau)\right],$$
$$\frac{dA_2}{dt} - \frac{i\Delta}{2}A_2 + \gamma A_2 = \alpha_2 e^{i\theta}\left[(1-k)\left(1-|A_2(t-\tau)|^2\right)A_2(t-\tau) + k\left(1-|A_1(t-\tau)|^2\right)A_1(t-\tau)\right].$$
(17)

Fig. 5(a) shows a chart of dynamical regimes on the $\Delta - k$ parameter plane for $\alpha_{1,2} = 2.2$, when the isolated oscillators operate in a single-frequency regime. If we introduce the coupling between the oscillators, a beating regime (i.e. two-frequency quasiperiodic oscillation) occurs. With an increase in the coupling parameter $k$ the beating regime is replaced by the synchronization regime. For small values of $\Delta$ synchronization occurs as a result of mutual frequency locking. In Fig. 6, typical spectra and waveforms are presented. With an increase in $k$ the fundamental frequencies of the first and the second oscillator shift towards one another, and the number of sidebands increases. The amplitudes $R_{1,2}$ periodically oscillate in time, and the oscillation period increases with $k$ (Figs. 6(a)-(d)). Finally, the fundamental frequencies of the oscillators coincide and the single-frequency regime with constant amplitudes establishes, as shown in Fig. 6(e).

At high values of the frequency mismatch, the synchronization occurs as a result of the suppression of natural frequency of one of the oscillators. Typical spectra and waveforms are presented in Fig. 7. With an increase in $k$ the spectral component at the natural frequency of the first oscillator decreases (Fig. 7(a),(b)) and vanishes (Fig. 7(c)). At $k = 0.47$ the mode of synchronization with constant amplitudes arises. Further increase in $k$ results in gradual decrease in the amplitudes $R_{1,2}$. Note that the amplitude of one oscillator significantly exceeds the amplitude of another one. At higher values of $k$ both amplitudes become zero (Fig. 7(d)), i.e. the amplitude death (AD) occurs, which is typical for systems with dissipative coupling [22-29]. Thus, the synchronization domain is located in a narrow zone between the beating and the AD domains. This regime known as broadband synchronization (BS) has previously been studied in [4,37,38]. Usually BS takes place in systems with non-identical excitation parameters which determine the oscillation amplitudes. However, in this case BS occurs even when the excitation parameters are identical ($\alpha_1 = \alpha_2 = \alpha$). This is due to the fact that variation of the frequency mismatch leads to the variation of the position of eigenfrequencies relative to the center of the system passband. Correspondingly, the oscillation amplitude also varies with $\Delta$. Since a time-delayed system with infinite-dimensional phase space has an infinite number of eigenmodes [3,17,18,31,35,36], the eigenfrequencies of different modes are alternately in resonance with the central frequency for different values of $\Delta$. Therefore, the BS domain looks not like a horizontal band as in [4,37,38], but like a periodical wavy-shaped band. At some points it touches the $\Delta$-axis in Fig. 5(a). At these points one of the oscillators is close to the boundary of a generation zone and its oscillation is suppressed. Here we can draw an analogy with [12,18], where the synchronization domain looks like the set of narrow tongues located at different eigenfrequencies. In such a way, the multimode nature of systems with delay becomes apparent.

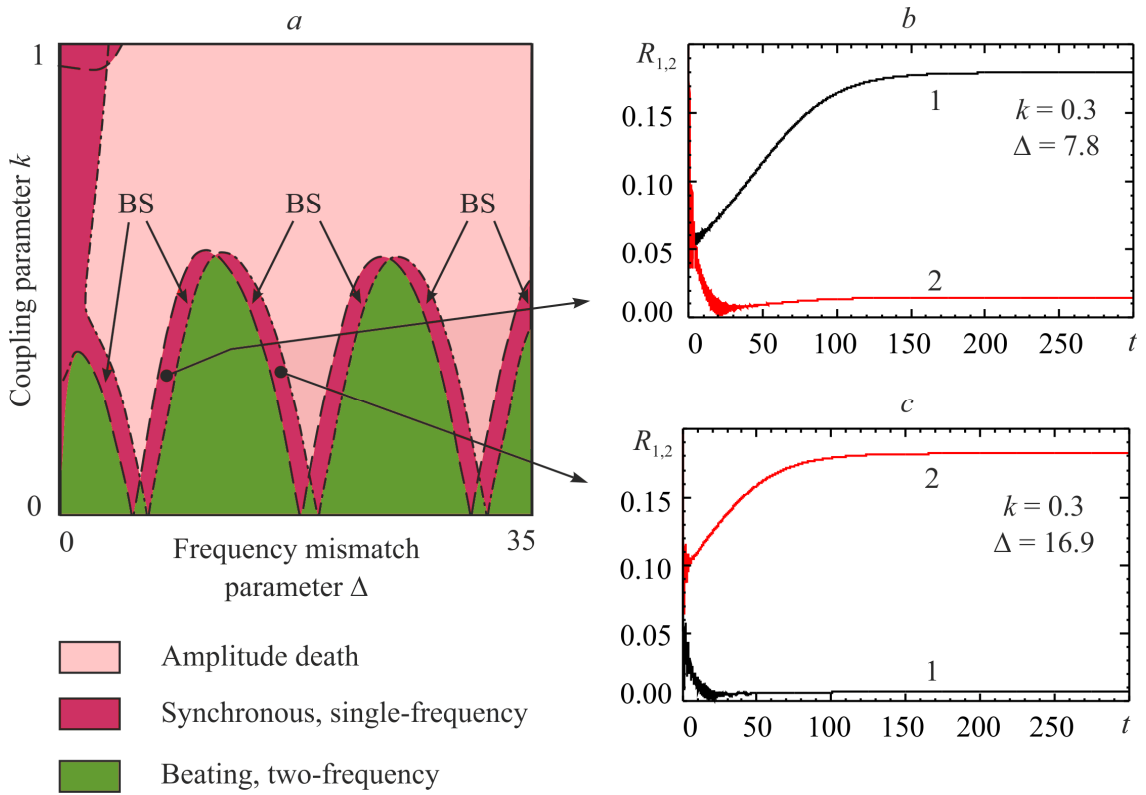

**Fig. 5.** (a) Chart of dynamical regimes for the system (17) for $\tau = 1.0$, $\gamma = 1.0$, $\theta = 0.1\pi$, $\alpha_{1,2} = 2.2$ and (b, c) waveforms at two points inside the BS domain. The analytical stability boundaries (20) are shown by the dashed and dashed-and-dotted lines.

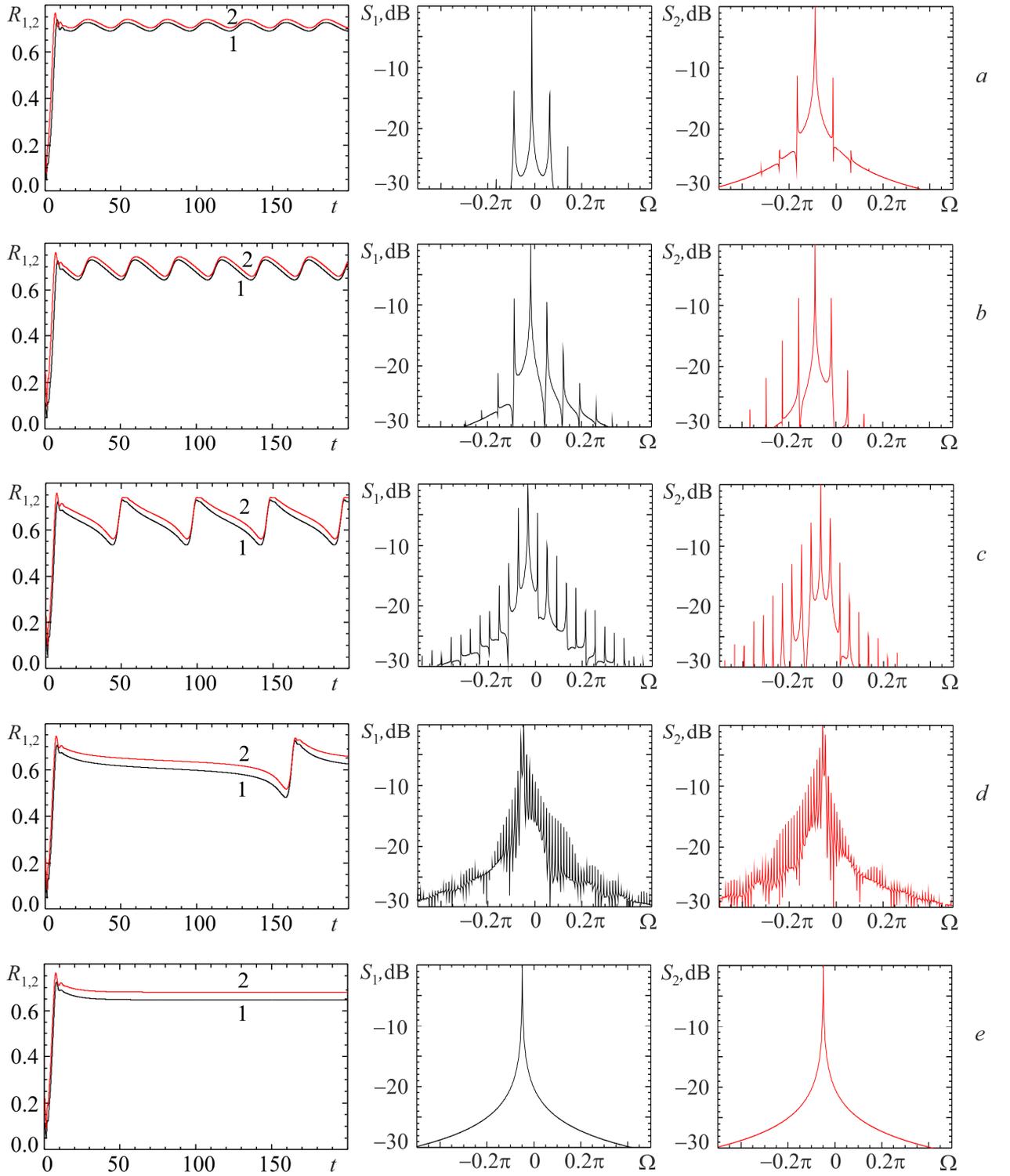

**Fig. 6.** Waveforms (left) and spectra for the first (center) and the second (right) oscillator demonstrating mutual frequency locking for the system of coupled oscillators (17). Values of the parameters are $\tau = 1.0$, $\gamma = 1.0$, $\theta = 0.1\pi$, $\alpha_{1,2} = 2.2$, $\Delta = 0.5$, (a) $k = 0.05$, (b) $k = 0.1$, (c) $k = 0.17$, (d) $k = 0.19$, (e) $k = 0.2$.

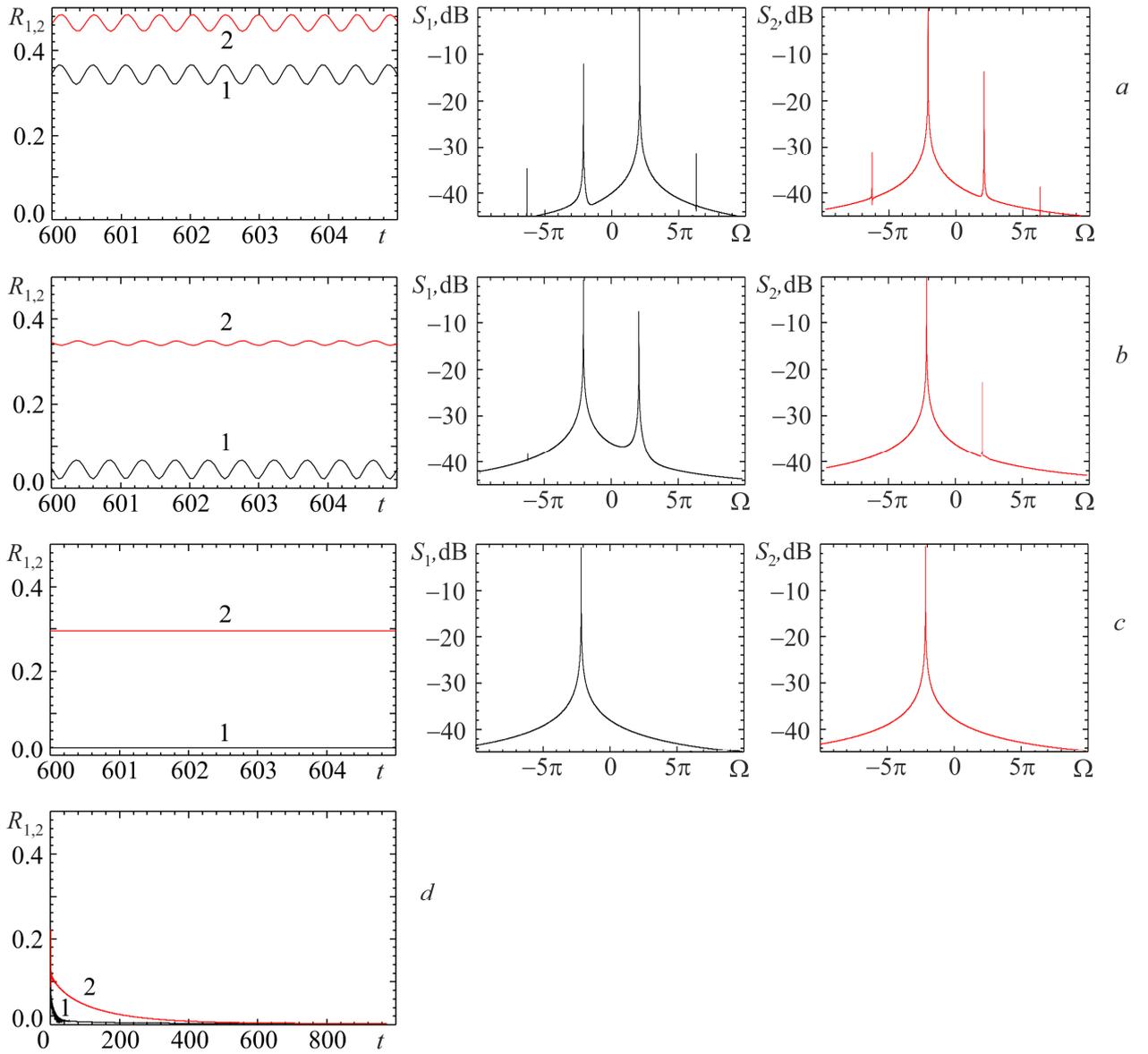

**Fig. 7.** Waveforms (left) and spectra for the first (center) and the second (right) oscillators demonstrating suppression of eigenmodes of one of the oscillators in the system (17). Values of the parameters are $\tau = 1.0$, $\gamma = 1.0$, $\theta = 0.1\pi$, $\alpha_{1,2} = 2.2$, $\Delta = 14.0$, (a) $k = 0.4$, (b) $k = 0.464$, (c) $k = 0.48$, (d) $k = 0.53$.

Inside the BS domain, natural oscillation of one oscillator is suppressed and another oscillator excites oscillation in the whole system. This fact is evidenced by a different level of oscillation amplitudes in Figs. 5(b),(c) where one of the amplitudes is much smaller than the other one. With an increase in the frequency mismatch $\Delta$ the effect of alternation of the leading oscillator [37] is observed. Namely, the first oscillator dominates inside the BS domain to the left of the beating domain (Fig. 5(b)), while the second oscillator dominates to the right of the beating domain (Fig. 5(c)). In contrast with systems with low number of degrees of freedom [37], we observe multiple alternation of the leading oscillator with variation of $\Delta$.

The boundaries for the AD and BS domains can be obtained analytically as stability condition for zero solution $A_{1,2} = 0$. Linearizing Eqs. (17) and seeking for a solution $A_{1,2} = R_{1,2} \exp(i\omega t)$, we obtain

$$\left(i\omega+\frac{i\Delta}{2}+\gamma-\alpha_1(1-k)e^{i\psi}\right)R_1 = \alpha_1 k R_2 e^{i\psi},$$
$$\left(i\omega-\frac{i\Delta}{2}+\gamma-\alpha_2(1-k)e^{i\psi}\right)R_2 = \alpha_2 k R_1 e^{i\psi}, \tag{18}$$

where $\psi = \theta - \omega\tau$. At the stability boundary the oscillation frequency $\omega$ is real.

Multiplying Eqs. (18) and separating real and imaginary parts of the obtained equation gives

$$\frac{\Delta^2}{4}+\gamma^2-\omega^2+(k-1)(\alpha_1+\alpha_2)\gamma\cos\psi+(1-2k)\alpha_1\alpha_2\cos 2\psi +$$
$$+\frac{k-1}{2}\bigl[(\alpha_1-\alpha_2)\Delta - 2(\alpha_1+\alpha_2)\omega\bigr]\sin\psi = 0,$$
$$2\gamma\omega - \frac{k-1}{2}\bigl[(\alpha_1-\alpha_2)\Delta - 2(\alpha_1+\alpha_2)\omega\bigr]\cos\psi + (k-1)(\alpha_1+\alpha_2)\gamma\sin\psi +$$
$$+(1-2k)\alpha_1\alpha_2\sin 2\psi = 0. \tag{19}$$

Solving Eqs. (19) numerically, one can plot the stability boundaries on the $\Delta-k$ plane.

For the oscillators with identical excitation parameters $\alpha_{1,2} = \alpha$, Eqs. (19) may be simplified as

$$k(\omega) = \frac{2\omega\gamma - 2\alpha(\gamma\sin\psi + \omega\cos\psi) + \alpha^2\sin 2\psi}{2\alpha^2\sin 2\psi - 2\alpha(\gamma\sin\psi + \omega\cos\psi)},$$
$$\Delta(\omega) = 2\sqrt{\alpha^2 k^2\cos 2\psi + \bigl[\omega-(1-k)\alpha\sin\psi\bigr]^2 - \bigl[\gamma-(1-k)\alpha\cos\psi\bigr]^2}. \tag{20}$$

Eqs. (20) give the stability boundaries written in parametric form. These boundaries are in excellent agreement with the numerical results, as shown in Fig. 5. They represent two periodically intertwined curves, which correspond to the suppression of oscillations in the first ($\omega<0$) and the second ($\omega>0$) oscillator, shown by the dashed and dashed-and-dotted lines, respectively. The BS regime is located between these two curves, while the AD domain is located above them. Thus, Eqs. (19), (20) describe not only the AD boundaries, but also the BS boundaries, except the region of small $\Delta$, where synchronization occurs as a result of the frequency locking, instead of suppression.

Now consider the case $\alpha_{1,2} = 3.2$ when, depending on the frequency mismatch, the isolated oscillators can operate either in self-modulation or in single-frequency regimes. Thus, introduction of small coupling leads to the two-frequency or three-frequency quasiperiodic oscillations, respectively. Fig. 8 shows the corresponding chart of dynamical regimes. A transition from the three-frequency to the two-frequency oscillation occurs as a result of suppression of the self-modulation frequency. With further increase in $k$ a transition to BS and then to AD occurs as described above. For small values of $\Delta$ a mutual locking of fundamental frequencies is observed, and one more domain of two-frequency quasiperiodic regime appears. In this domain, the spectrum differs from the spectrum of the beating regime since it contains one fundamental frequency and symmetrically spaced self-modulation sidebands. With an increase in $k$ transition to synchronization regime via suppression of the self-modulation frequency occurs. However, at high values of $k$ self-modulation domain arises again and then transition to chaos occurs.

A special character of the frequency locking which takes place at small values of $\Delta$ and $k$ should be noted. Fig. 9 shows the waveforms and spectra for one of the oscillators. For the other one they look similar and are not shown in Fig. 9. At $\Delta = 0.7$ two-frequency beating regime is observed (Fig. 9(a)). With the decrease in $\Delta$ locking of the fundamental frequencies occurs and the beating period increases. However, on the contrary to Fig. 6, the spectral components at frequencies which are close to the frequency of self-modulation satellites have large amplitudes. In the beating regime average amplitude oscillates in time. At the moments when this amplitude increases, excitation of the self-modulation satellites occurs and appearance of high-frequency modulation of the waveform is clearly visible. When the average amplitude decreases, the satellites decay. Such a waveform resembles spike-burst oscillation typical for neural dynamical systems [14]. With the decrease in $\Delta$ the beating period increases, but the self-modulation frequency remains constant (Figs. 9(b),(c)). At $\Delta = 0.55$ locking of the fundamental frequencies occurs and, simultaneously, the self-modulation disappears (Fig. 9(d)). We notice that such an unusual kind of frequency locking has been observed in previous studies of the forced synchronization of a delayed-feedback oscillator [12, 31]. Such a behavior should be common for synchronization of systems with self-modulated oscillations, which are typical for many microwave electronic oscillators. With further decrease in the frequency mismatch the self-modulation arises again, as shown in Fig. 9(e). The oscillators remain synchronized but self-modulation sidebands appear on both sides of the fundamental frequency.

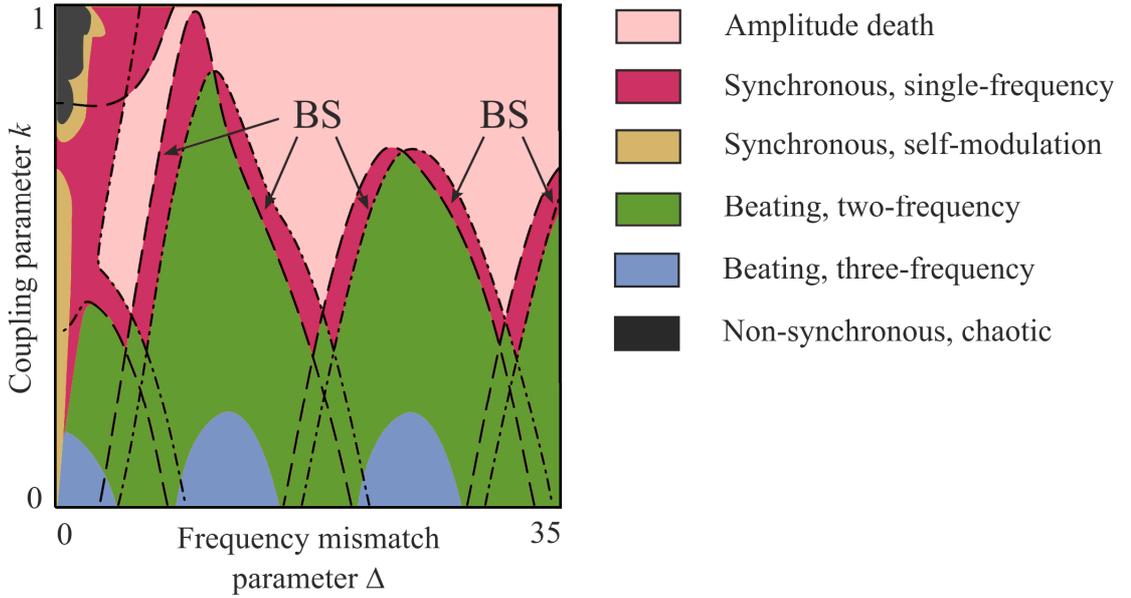

**Fig. 8.** Chart of dynamical regimes for the system (17) for $\tau = 1.0$, $\gamma = 1.0$, $\theta = 0.1\pi$, $\alpha_{1,2} = 3.2$. Dashed and dashed-and-dotted lines show analytical stability boundaries (20).

Fig. 10 shows charts of dynamical regimes on the $\Delta - k$ parameter plane for the oscillators with non-identical excitation parameters. For the first oscillator we choose $\alpha_1 = 3.2$, i.e., the isolated oscillator operates in the self-modulation regime. For the second oscillator the excitation parameter is chosen in the domain of single-frequency regime $\alpha_2 = 1.2$ (Fig. 10(a)) and $\alpha_2 = 2.2$ (Fig. 10(b)). Domains of synchronous regimes, BS, AD, two- and three-frequency beating regimes are clearly visible on the charts. The mechanisms of transition between different regimes are the same as in the case of identical $\alpha_{1,2}$. One can see that the larger the difference of the excitation parameters, the wider the BS domain. In contrast to Figs. 5,8, the boundary of suppression of the first oscillator (dashed lines) is located above the boundary of suppression of the second one (dashed-and-dotted lines). Thus, the oscillator with larger excitation parameter always dominates (Fig. 10(c),(d)) and the effect of alternation of the leading oscillator is not observed.

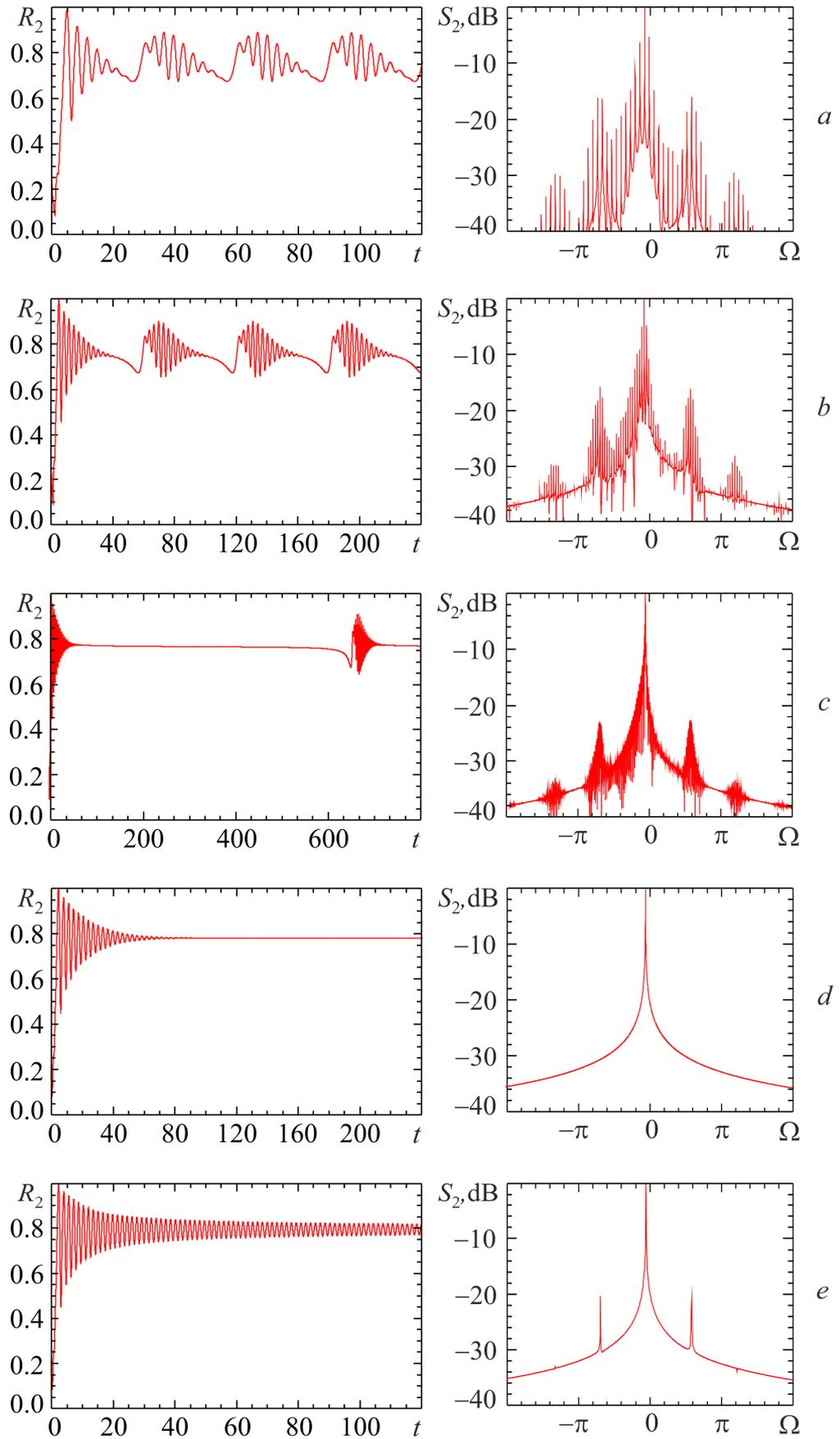

**Fig. 9.** Waveforms (left) and spectra (right) for the second oscillator in the system (7). Values of the parameters are $\tau = 1.0$, $\gamma = 1.0$, $\theta = 0.1\pi$, $\alpha_{1,2} = 3.2$, $k = 0.21$, (a) $\Delta = 0.7$, (b) $\Delta = 0.6$, (c) $\Delta = 0.565$, (d) $\Delta = 0.55$, (e) $\Delta = 0.519$.

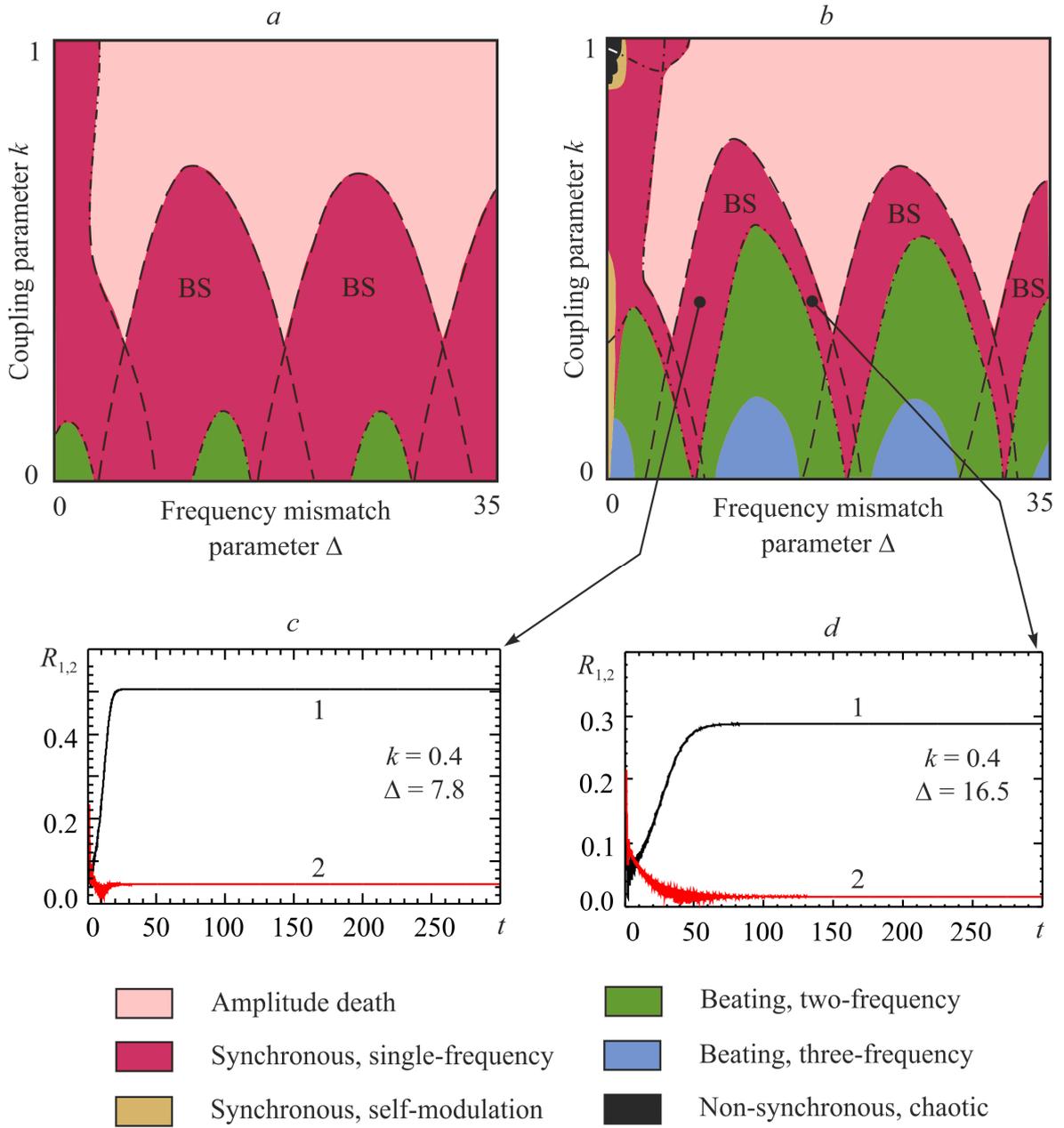

**Fig. 10.** Charts of dynamical regimes for the system (17) for $\tau=1.0$, $\gamma=1.0$, $\theta=0.1\pi$, $\alpha_1=3.2$, (a) $\alpha_2=1.2$, (b) $\alpha_2=2.2$ and (c,d) waveforms at two points inside the BS domain.. Dashed and dashed-and-dotted lines show analytical stability boundaries (19).

## 5. Conclusion

The main objective of this paper is to study mutual synchronization between two time-delayed oscillators, which are coupled by feeding a power portion from the feedback loop of one oscillator into the feedback loop of its counterpart, and vice versa. The model of two coupled oscillators consisting of the nonlinear power amplifier, the bandpass filters, and the delayed-feedback loop is considered. The system of DDEs (7) with a special type of coupling, namely, with nonlinear time-delayed dissipative coupling, is derived. The model studied in this paper describes synchronization of two electronic microwave oscillators, surface acoustic wave oscillators, or ring-loop optical laser oscillators.

In the system of two coupled oscillators with identical parameters, the complete synchronization regimes with equal amplitudes, i.e. $A_1=A_2$, are observed. The oscillators may synchronize in the single-frequency, self-modulation, and chaotic regimes. In case of complete chaotic synchronization, transitions between synchronous and asynchronous modes occur through the intermittency of "bubbling" type.

The mechanisms of transition from beating to synchronous regime are studied for oscillators with frequency mismatch. The cases of coupling of two oscillators operating in a single-frequency regime, as well as in a self-modulation regime, are considered. In the parameter plane "frequency mismatch $\Delta$ — coupling parameter $k$", besides the main synchronization tongue located in the domain of small $\Delta$, there appears a BS domain which looks like a band extending to very large frequency mismatches. Inside this band, one of the oscillators dominates and suppresses natural oscillation of its counterpart. In contrast with previously studied systems with low-dimensional phase space, BS takes place even when the excitation parameters are identical. Numerical simulation reveals the complicated wavy shape of the BS domain which is nearly periodic in $\Delta$. This occurs due to the fact that variation of $\Delta$ leads to the variation of the eigenfrequency and amplitude of the time-delayed oscillator. Thus, when the frequency mismatch is varied, different modes of the multi-mode time-delayed oscillator are alternately in resonance with the central frequency of the filter, and either one or the other oscillator dominates. Analytical formulas for the boundaries of AD and BS domains are obtained. They completely agree with the results of numerical simulation.

If the oscillators have non-identical excitation parameters, the BS domain becomes wider with the increase in the non-identity. If the excitation parameter of one oscillator is much larger than that of another one, the former oscilator is always dominant and the effect of alternation of the leading oscillator is not observed.

Time-delayed systems have infinite dimensional phase space and can operate in various regimes including chaotic ones. Thus, the picture of synchronization of such oscillators reveals many features caused by the effect of delay and existence of multiple eigenmodes.

**Acknowledgments**

This work is supported by the Russian Foundation for Basic Research grant No 12-02-31493 and 14-02-31410.